## 1.1 Observations and Open Questions in Beam-Beam Interactions

Tanaji Sen
Accelerator Physics Center, FNAL, Batavia, IL 60510, USA
Mail to: tsen@fnal.gov

### 1.1.1 Introduction

The first of the hadron colliders, ISR, started operation in 1970. In the following years, the hadron colliders to follow were the SPS (started 1980), the Tevatron (started 1987 first as a fixed target machine), RHIC (started 2000) and most recently the LHC, which started in 2008. HERA was a hybrid that collided electrons and protons. All of these accelerators had or have their performance limited by the effects of the beam-beam interactions. That has also been true for the electron-positron colliders such as LEP, CESR, KEKB and PEPII. In this article I will discuss how the beam-beam limitations arose in some of these machines. The discussion will be focused on common themes that span the different colliders. I will mostly discuss the hadron colliders but sometimes discuss the lepton colliders where relevant. Only a handful of common accelerator physics topics are chosen here, the list is not meant to be exhaustive. A comparative review of beam-beam performance in the ISR, SPS and Tevatron (ca 1989) can be found in reference [1].

Table 1 shows the relevant parameters of colliders (excluding the LHC), which have accelerated protons.

Table 1: Basic parameters of past and present fully commissioned hadron colliders

|  | ISR | SPS | Tevatron | HERA p | RHIC |
|---|---|---|---|---|---|
| Circumference [m] | 943 | 6911 | 6283 | 6336 | 3834 |
| Energy [GeV] | 31 | 315 | 980 | 920 | 250 / 100 |
| Peak Luminosity [$\times 10^{32}$ cm$^{-2}$ s$^{-1}$] | 1.3 | 0.06 | 4.0 | 0.5 | 0.85 |
| Lumi lifetime [hrs] | ? | 9 | 6 | ? | 6 |
| # of head-on collisions | 8 | 3 | 2 | 2 | 2 |
| Number of parasitics | 0 | 9 | 70 | 0 | 4 |
| Total bm-bm spread | 0.008 | 0.015 | 0.025 | 0.003 | 0.013/0.009 |
| $\beta_x^*, \beta_y^*$ [m] | 30, 0.3 | 0.6, 0.15 | 0.28, 0.28 | 2.45, 0.18 | 0.7 / 0.7 |
| $\varepsilon_x, \varepsilon_y$ [rms, π μm] |  | 2.75, 2.75 3, 2.5 (a) | 2.9, 3.3 1.6, 1.4(a) | 3.7, 3.7 | 3.3, 3.3 |
| Bunch intensity [$\times 10^{11}$] | - | 1.3(p) 0.7(a) | 3.1 (p) 1(a) | 0.9 | 1.3 |
| Number of bunches | N/A | 6 | 36 | 180 | 110 |
| Bunch spacing [nsec] | N/A | 1150 | 396 | 96 | 108 |
| Bunch length [m] | N/A | 0.72 | 0.6 | 0.30 | 0.6 / 0.8 |

Notation: a = anti-protons

Luminosity lifetime in the table refers to the initial luminosity lifetime at the start of stores. ISR also collided protons-antiprotons but peak luminosities were reached with protons in both beams. HERA was an e-p collider but is included here.

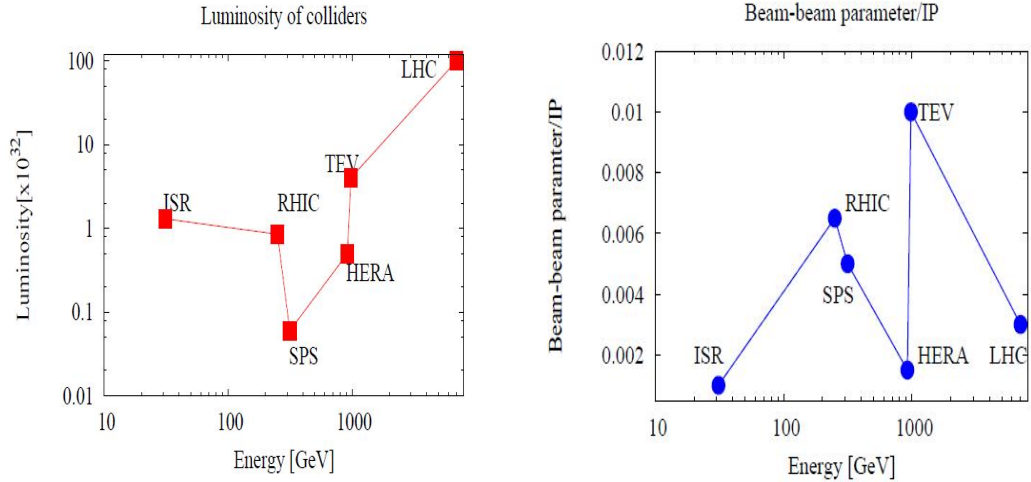

**Figure 1**: Left plot shows the luminosity in cm$^{-2}$s$^{-1}$ vs. beam energy. Right plot shows the beam-beam parameter per IP vs. beam energy. For SPS and the Tevatron, the beam-beam parameter for the anti-protons is shown. Also shown are the parameters for the LHC at its design energy and luminosity.

Figure 1 shows the luminosity and beam-beam parameter/IP $\xi$ for the different colliders. While the SPS had the lowest luminosity (because of the fewest number of bunches), it had small emittance bunches and had the highest specific luminosity so far.

Tune space: In the Tevatron, the working points lie above the half integer between the 5$^{th}$ and 7$^{th}$ order resonances with an available tune space of 0.028 which is comparable to the total beam-beam tune spread. In RHIC, the working points also above the half integer lie between 3$^{rd}$ and 10$^{th}$ order resonances with an available tune space of 0.03. The maximum tune spread is about half this value. The SPS also operated within these resonances. In HERA-p, tunes were below the half integer but placed between 7$^{th}$ and 10$^{th}$ order resonances with an available tune space of 0.014, several times the beam-beam induced tune spread for protons. In most of these colliders, the tunes have to be controlled to within 0.002 for optimal operation. This is not always easy, e.g. in the Tevatron the proton tune spread is determined by the anti-proton bunch intensity which can vary significantly from bunch to bunch.

### 1.1.2 Beam-beam limits in different colliders

Limits imposed by the beam-beam interactions can manifest in several different ways. Here we briefly review how the limits arise/arose in different colliders.

### *1.1.2.1    Hadron colliders*

Tevatron

Beam-beam interactions impose limits at all stages of the operation cycle and in different ways. At injection, the limits are imposed by the long-interactions when the two beams with 36 bunches each circulate on their helical orbits and each bunch suffers 72 long-range interactions around the ring. Both beams suffer losses proportional to the intensity of the other beam. At collision with 2 head-on interactions and 70 long-range interactions, the limiting processes are different for the two beams. The long-range interactions contribute a tune spread of about 0.008, equal to from each of the main collisions. In current operations, both species have about the same beam-beam tune spread and are effectively in the strong-strong regime. Early in Run II, anti-protons suffered large losses during the beta squeeze and stores due to the long-range interactions, particularly the 4 interactions with smallest separations on either side of the 2 IPs. In 2006, additional separators were installed to increase the beam separations from about 5.4σ to about 6 σ [2] at these locations. Beginning in 2005 electron cooling of anti-protons in the Recycler was made operational which made their emittance much smaller than those of protons [3]. Consequently the anti-protons effectively experience only the linear part of the head-on beam-beam force and do not suffer much from it. Since 2006, anti-proton losses due to beam-beam interactions during stores have been small, provided the tunes are well controlled. Protons on the other hand have tunes closer to $12^{th}$ order resonances and are transversely larger than the anti-protons. Consequently during head-on collisions, they experience the non-linear beam-beam force enhanced by chromatic effects and suffer beam loss and emittance growth. Long-range interactions have affected protons occasionally during the beta squeeze when separations can drop to low values.

Earlier reports on beam-beam phenomena early in Run II can be found in several references, e.g. [4, 5, 6]. A review of beam-beam observations in Run I can be found in [1]. In 2010 the Tevatron achieved a peak luminosity of $4 \times 10^{32}$ cm$^{-2}$ s$^{-1}$, about three times the peak value obtained with the ISR. Summaries of recent improvements made to the Tevatron complex can be found in [3, 4].

RHIC

RHIC has collided many species including proton-proton, gold-gold, gold-deuteron and others. I will discuss here some of the limits observed with proton-proton collisions. The beams in the Blue and Yellow rings have nearly the same intensity and emittances, so RHIC operates in the strong-strong regime, as does the LHC. During injections and acceleration, the beams are have a large enough vertical separation that long-range interactions at 4 locations do not lead to any losses. At collision each bunch suffers 2 head-on collisions. During recent runs, the beam-beam parameter $\xi$ per IP has approached 0.0065, close to the value in the Tevatron [8]. Dominant sources of beam lifetime limitations, not due to luminosity burn up, include beam-beam effects, IR multipole errors and parametric modulations due to mechanical vibrations of the triplets [9]. RHIC operates between the $3^{rd}$ and $10^{th}$ order resonances. When the tunes get too close to the $10^{th}$ order resonances, both luminosity lifetime and proton polarization (which may be affected by beam-beam) suffer. During the latest runs, beta* values became comparable to the bunch length and the hourglass effect became significant enough to reduce the luminosity [10]. At intensities beyond $2 \times 10^{11}$ /bunch, the beam-beam tune spread will exceed the resonance free space. There are plans to use electron lenses to compensate the effects of these head-on interactions.

### HERA

The beam-beam parameter for HERA-p was almost a factor of 10 lower than in the Tevatron. Also, as remarked above, the resonance free space was about 4-5 times the beam tune spread. Nevertheless the head-on interactions did induce beam losses.

During the early commissioning stage, proton transverse beam sizes were about 3-4 times the electron beam sizes and their lifetime during stores was very low, around 0.5 hours. As the proton beam size was reduced to match the electron size, the lifetime improved to about 100 hours or more [11]. During 2003 and 2004, proton beams were observed to be driven by coherent oscillations of the lepton beam when the tunes of the two beams approached resonances too closely. Under extreme conditions, the proton beam emittance grew by a factor of 2-4 times [12]. This growth was avoided by careful choice of the tunes and by bringing the beams into collision sequentially at the two IPs. In the final years of operation, increasing beam-beam forces on the protons increased diffusion into the beam halo and background rates and thus led to a "soft limit" rather than a hard limit [13]. However orbit vibrations at the IP due to mechanical vibrations of the triplet by more than a few microns were considered intolerable.

The lepton beam-beam limit in HERA was primarily due to operation close to the integer tune in order to maximize polarization. When the beam-beam tune spread overlapped low order synchro-betatron resonances, coherent oscillations and emittance growth of the lepton beam resulted. Careful control of the tunes was necessary to avoid these resonances [13].

### SPS

Prior to 1988 the SPS operated with 3 proton bunches and 3 anti-proton bunches circulating in the same vacuum chamber. The protons had an emittance about 4 times larger than that of the anti-protons. During the start of stores, proton loss rate was high with an initial lifetime of around 10 hours and the background rates were unacceptably large [14]. Protons in the transverse tails were sensitive to very high order resonances such as the 16[th] order and were lost. The losses were controlled by a controlled increase of the anti-proton emittance at the start of the stores – similar to what is done now in the Tevatron. Along with other upgrades in 1988, the proton to anti-proton emittance ratio was reduced to 12/7 and the number of bunches in each beam was increased to 6. During injection and acceleration, the beams were horizontally separated with electrostatic separators. At injection, the beam separations at the 12 parasitic interactions varied between 1.3 to 7.9 units of the anti-proton beam size [14]. Beam losses due to 7[th] order resonances were associated with these interactions during injection and acceleration. At top energy each bunch had 3 head-on collisions, two at the experiments and one in between them. During stores with more equal beam sizes, the protons were now sensitive to lower order resonances such as 10[th] order but background rates were acceptable and the initial proton lifetime had increased to about 50 hours. A comparative review of SPS and Tevatron performance up to 1989 can be found in [15].

## ISR

This machine had two interleaved rings in which first unbunched beams of protons and later antiprotons and other particles were brought into collision. There were 8 crossing points of which 5-6 were used for experiments [16, 17]. Another feature of the ISR was that it had a working line, shown in Figure 2, rather than a working point.

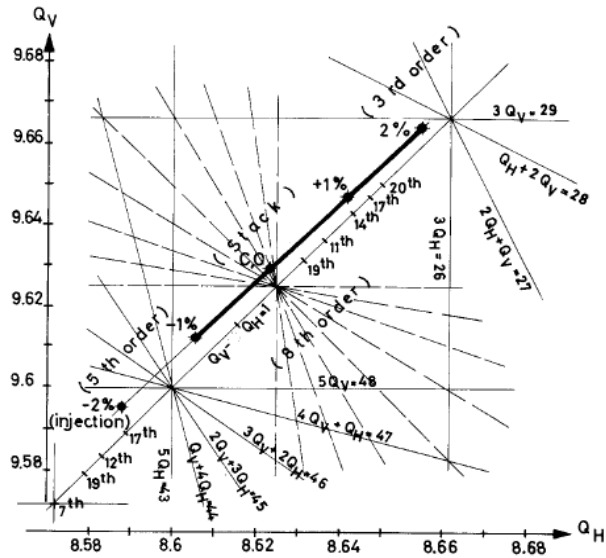

**Figure 2**: One of the working lines in the ISR (named 8C) between 3$^{rd}$ and 5$^{th}$ order resonances and straddling 8$^{th}$ order resonances (taken from [18])

This large tune spread was required for stability against the transverse resistive wall instability. As a consequence, the beams crossed some low order betatron resonances which led to particle loss. Synchro-betatron resonances were not an issue. During collisions beam-beam effects also led to particle loss, often from coherent effects. This will be discussed further below. Beam currents in the range of 30-40 Amps were stored during high luminosity runs with lifetimes in the tens of hours. Overviews of the accelerator physics issues in the ISR can be found in [19, 20].

### 1.1.2.2 $e^+$-$e^-$ colliders

## KEKB

Prior to 2007, beams in KEKB had a crossing angle of 22 mrad at the IP. Crab cavities were introduced in 2007, one in each ring, to have effective head-on collisions and recover the geometric loss of luminosity. However when the bunch currents were raised beyond values circulated without the crab cavities, beam lifetimes dropped. The lifetimes could be improved by introducing horizontal offsets in the crab cavities, the amount of offset depended on the bunch current. In 2008 it was understood to be due to the dynamic beta beating from the beam-beam interaction and operation close to a half integer [21]. The horizontal beam sizes of the beams were large at the crab cavities that did not have sufficient aperture. The optics was changed to reduce $\beta_x$ at the cavities, $\beta^*$ was raised to 0.15m to improve lifetime. The most important improvements came from the installation of skew sextupoles around the IR to reduce chromatic coupling at the IP.

These alone raised the luminosity by 15% and led to a peak luminosity of $2.1 \times 10^{34}$ cm$^{-2}$s$^{-1}$ in 2009 [22, 23].

PEPII

During 2008, its last year of operation, PEPII operated with 1732 bunches in each ring and achieved a peak luminosity of $1.21 \times 10^{34}$ cm$^{-2}$s$^{-1}$ [24]. At the highest bunch currents, the performance was limited by the head-on beam-beam interactions. For example, the low energy ring (LER) currents were limited by the losses and backgrounds from the beam in the high energy ring (HER). Additionally increasing the beam current in the LER also increased its own beam size, which was not understood [24]. The maximum beam-beam parameter achieved was 0.113 in the horizontal plane of the HER.

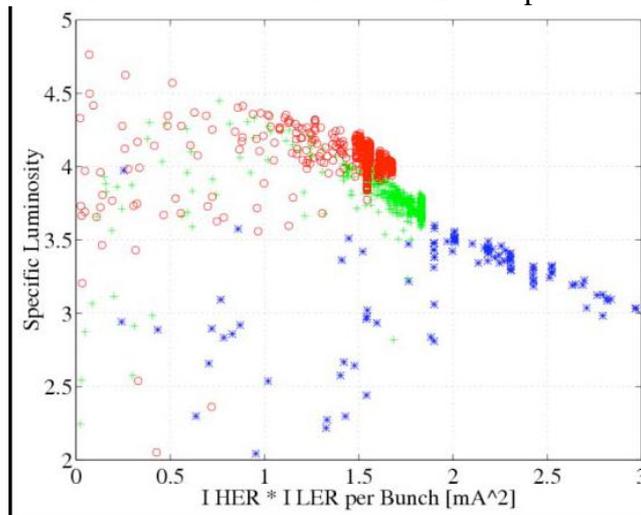

**Figure 3**: The specific luminosity vs. the product of the bunch currents in the two rings in PEP II (taken from reference [24]). At low currents the dynamic beta effect increased the luminosity by decreasing beam sizes at the IP but at higher currents, losses due to the beam-beam interactions reduced the specific luminosity.

The effect of the parasitic collisions on the luminosity was reduced to a few percent, after correcting for the tune shift and coupling generated by the vertical separation of the beams at these locations. In the early years of operation, electron cloud effects in the LER had to be mitigated by solenoidal fields in the straight sections, addition of antechamber, photon stops and TiN coatings in the arcs. A complete list of improvements made to PEPII over the years can be found in reference [24].

CESR

Until 2001 CESR operated as a symmetric energy collider at 5 GeV with electrons and positrons circulating in the same beam pipe. In 2001 there were nine bunch trains in each beam with 4 bunches per train for a total of 71 long-range interactions and 1 head-on collision. The beams were horizontally separated into pretzel orbits by electrostatic separators. The beam separations appear to have ranged from 4 to 7 $\sigma$ [25]. The bunch currents were limited by the parasitic interactions. When a 5$^{th}$ bunch was added to each train, the specific luminosity and the beam lifetimes suffered [25]. Attempts to increase the bunch current beyond 7.5 mA with 4 bunches in each train also led to lower lifetimes. The average beam-beam tune shift in the vertical plane saturated at 0.07. In 2001 CESR became CESR-c to study the bound states of charmed quarks and the energy was lowered to 2 GeV. During 2006 it operated with 8 trains of 3 bunches each, so each bunch suffered 47 long-range interactions and 1 head-on collision. Beam-beam effects were more severe at the lower energy. After local compensation of the phase advance shifts and beta-beats due to the long-range interactions, bunch currents could be raised to 3 mA from 2.5 mA before the compensation [26].

### 1.1.3 Scaling laws

Scaling laws which relate how beam loss rates or luminosity lifetimes relate to beam parameters can be useful for predicting the changes when beam parameters change in a given machine, for example after an upgrade. However these laws depend on the details of the machine and can usually not be applied across different accelerators. Furthermore even in a single machine, it is hard to measure beam loss rates or emittance growth against a single variable (such as bunch intensity of the opposing beam) over a wide enough range and with enough statistics, keeping all other factors constant. This is usually due to the lack of dedicated study time. Typically the loss rates or beam growth are measured at different points in time when other machine parameters (such as orbits, tune, chromaticities etc) may have also changed. With these caveats in mind, we now take a look at some scaling laws, some of which were obtained from data taken during machine experiments.

#### *1.1.3.1  Tevatron: Losses at injection*

At injection, the long-range interactions are responsible for losses. In 2005, the losses of anti-protons and protons were fitted to some key parameters. Figure 4 shows the proton loss rate dependence on the horizontal chromaticity.

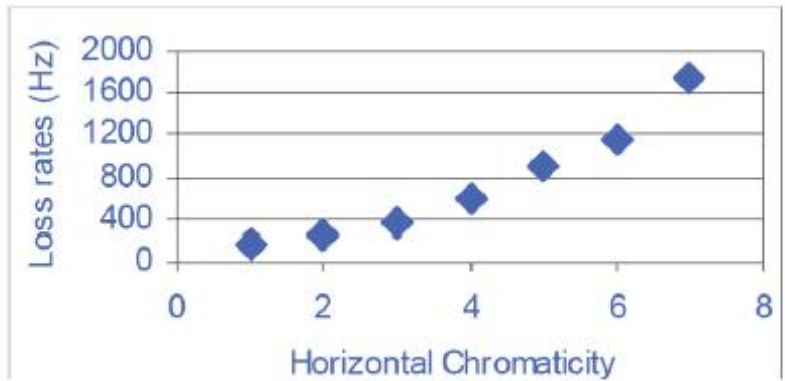

**Figure 4**: Dependence of proton loss rates in the Tevatron on horizontal chromaticity at injection

The empirical law relating proton and anti-proton losses to key parameters (adapted from [27]) was found to be

$$\frac{\Delta N_{a,p}}{N_{a,p}} \propto \sqrt{t} N_{p,a} \varepsilon_{a,p}^2 Q_{a,p}^{'2} \mid \Rightarrow (Q_{x,y}, d_{a-p}, \varepsilon_L, D_{apert} : const) \qquad (1)$$

Here *t* is the time spent at injection, $N_p$, $N_a$ are the proton and anti-proton bunch intensities, $\varepsilon_p$, $\varepsilon_a$ are the proton and anti-proton transverse emittances, $Q'$ is the chromaticity. The above dependencies hold only if the variables in parentheses are held constant. $Q_a$, $Q_p$ the tunes, $d_{a\text{-}p}$, the separation between the beams, $\varepsilon_L$ the longitudinal emittance and $D_{apert}$ is the distance to the physical aperture. The functional dependencies on these parameters ($N$, $\varepsilon$, $Q'$) can be completely different if any of the variables held constant, e.g. the tunes, change. The $\sqrt{t}$ dependence can be explained as the initial time dependence of a normal diffusion process [28], which at long times progresses to the more familiar exp(-t) decay for the intensity. It would be desirable to develop a theoretical model that explains the linear dependence on the opposing beam intensity and the quadratic dependence on its emittance and chromaticity but such a detailed understanding has not yet been developed.

### 1.1.3.2 Tevatron: Anti-proton losses during stores

Anti-proton loss rates are determined mostly by the long-range interactions. Data taken during 2004-2005 could be empirically fit to the law [27]

$$\frac{1}{\tau_a} = \frac{1}{N_a}\frac{dN_a}{dt} \propto N_p \frac{\varepsilon_a^2}{d_{a-p}^3} \Big| \Rightarrow (Q_{x,y}, Q'_{x,y}, \varepsilon_p, \varepsilon_L, M, D_{apert} : const) \qquad (2)$$

where *M* is the bunch number in the train and $d_{a\text{-}p}$ is an average distance between the beams or more precisely the scale of the helix size compared to a nominal helix. The dependence on the beam separation was measured by changing the size of the helix everywhere in the ring by a scale factor. It is worth noting that changing the helix also changes tunes, coupling and chromaticities so their effects on beam loss may also be present. As at injection, the losses depended linearly on the opposing beam intensity and quadratically on its own transverse emittance. It is possible that these dependencies on ($N$, $\varepsilon$) are nearly universal for a well tuned machine away from harmful resonances. One machine where this could be tested is the LHC which also has several long-range interactions per turn.

### 1.1.3.3 SPS study of proton losses

It is very likely that the inverse cube power law dependence on the beam separation is not universal but depends on the details of the beam and machine parameters. One example in the SPS is drawn from a study done with a single proton bunch interacting with two anti-proton bunches [29]. At two points the beams collided head-on, at two other points they were separated by 6-7σ of the anti-proton beam size. The loss rate and background rates were measured during two horizontal tune scans, one with full separation and the other with half their separations at the parasitic interaction locations. Figure 5 shows the decay rate on the left vertical scale and the background rate on the right vertical scale. There was a jump in the rates (by a factor of 2 – 3) for the halved separation only at the 13[th] and 16[th] order resonances, but not at the 10[th] order resonance.

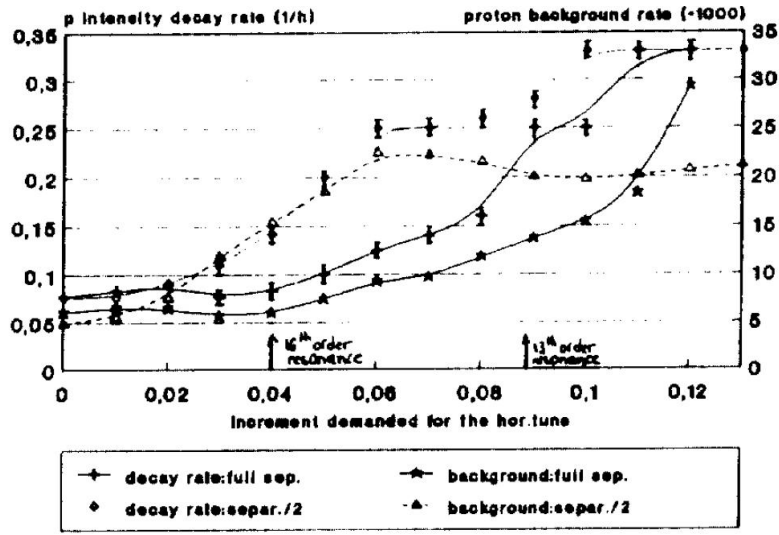

**Figures 5**: Proton intensity decay rate and proton background rate as a function of Ihe horizontal tune at two separations (taken from [29]).

The power law dependence on the separation is weaker in this measurement compared to the Tevatron data and it is tune dependent. The jump in rates at the 16[th] order resonance suggests that it was driven by the parasitic interactions but the 10[th] order resonance was not.

### 1.1.3.4 Tevatron: Proton losses due to head-on collisions

Proton loss rates during the first two hours of stores in 2008 are plotted against the product of the anti-proton bunch intensity and the ratio of emittances in Figure 6 below.

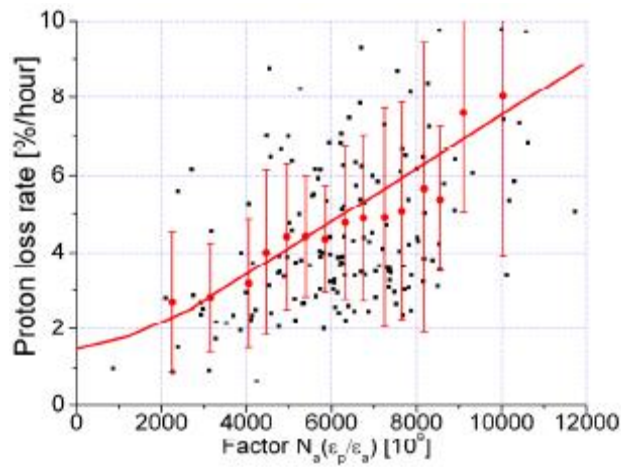

**Figure 6**: Proton loss rates vs. $N_a(\frac{\varepsilon_p}{\varepsilon_a})$ during stores in 2008 (taken from Ref [4])

This shows a nearly linear dependence of proton losses on this product. This suggests an empirical law

$$\frac{1}{\tau_p} = \frac{1}{N_p}\frac{dN_p}{dt} = N_a(\frac{\varepsilon_p}{\varepsilon_a}) \mid \Rightarrow (Q, Q', \beta^*, \varepsilon_L : const) \quad (3)$$

However note that the error bars on the data are fairly large. Also, there was not much variation in the proton emittance in this data.

The only models that exist to describe particle transport and beam loss in the absence of external noise are based on diffusion due to the overlapping of resonances. The diffusion coefficient is determined by the change in action which when dominated by beam-beam effects is proportional to the beam-beam parameter, hence

$$D(J) \sim \Delta J^2 \sim \xi^2 \quad (4)$$

Diffusion models therefore lead to diffusion coefficients that depend quadratically on the beam-beam parameter. In general extracting the lifetime from the diffusion coefficients requires solving a diffusion equation. In some cases the lifetime or loss rate can be extracted more directly. For example, the loss rate in the case of isotropic diffusion can be expressed in terms of the diffusion coefficients as [30]

$$\frac{1}{\tau} = N_D (\int \frac{J_r dJ_r}{D(J_r)})^{-1} \quad (5)$$

where $N_D$ is the number of dimensions (=2 or 3 if longitudinal effects are included), $J_r$ is the radial action and $D(J_r)$ is the radial isotropic diffusion coefficient. Thus the loss rate should also depend quadratically on the beam-beam parameter. The empirical fit above in Equation (3) shows a linear dependence on the beam-beam parameter. Reconciling theoretical models to the empirical fits remains a challenge.

**1.1.4    Influence of machine optics on beam-beam phenomena**

In all colliders global orbits, tunes, coupling, chromaticities etc have to be well controlled for optimum integrated luminosity. Here I will discuss some recent examples of how local optics parameters in the interaction regions and beam-beam interactions have influenced performance.

*1.1.4.1    Local and beam-beam chromaticity*

*Experience at the Tevatron*

Beam-beam effects can directly contribute to chromaticity. The head-on interactions can do so for bunches with lengths comparable to β* or for short bunches if the beams are not exactly round at the IP so that the beam-beam tune shift does depend on the β* values. Alternatively collisions at a crossing angle can also contribute to chromaticity. However these are usually relatively small contributions. Long-range interactions on the other hand have sextupole components in their multipole expansion and if these interactions occur at regions of non-zero dispersion can contribute significantly to the chromaticity. This is the case in the Tevatron where the contributions also differ bunch by bunch since each bunch has its own distribution of long-range separations and locations. The left plot in Figure 7 shows the theoretically calculated chromaticities at top energy due to the long-range interactions only, taken from [31]

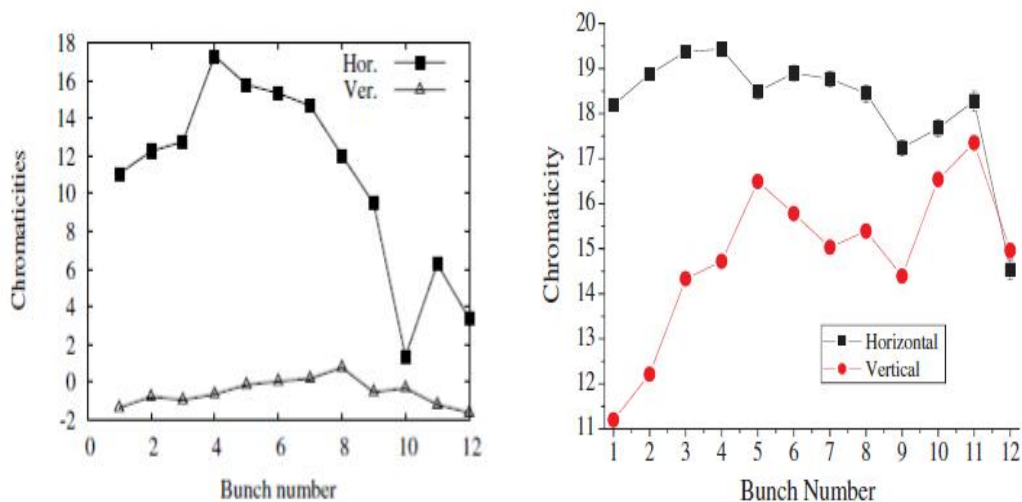

**Figure 7**: Left: Theoretical estimate of bunch-by-bunch chromaticity due to long-range interactions only. The right plot shows the measured bunch-by-bunch chromaticity that includes machine chromaticity and the effects of coupling as well.

In this theoretical calculation, the contributions to the vertical chromaticity are fairly small because the vertical dispersion is also small around the ring. However this does not take into account coupling between the two planes. The machine chromaticity, which would shift all the chromaticities by constant amounts, was not included. The measured bunch by bunch chromaticity in the Tevatron shown in the right plot of Figure 7, taken from [27], demonstrates (a) similar variation in chromaticity between the bunches and (b) that coupling tends to equalize the horizontal and vertical chromaticities.

It is worth noting that just like the tunes, the chromaticities also depend on the transverse amplitudes and chromaticity footprints exist which are also different for each bunch [32]. As with the beam-beam tune footprints, these footprints are hard to observe directly with measurements. However they can have observable consequences. If particles have chromatic tunes that lie near resonances, then their momentum deviation, their transverse amplitude and the specific bunch will determine which particles are lost due to these resonances.

The level of machine chromaticity also influences the effects of the long-range interactions. Prior to December 2008, the machine chromaticity in the Tevatron during the squeeze was kept between 12-14 units to stabilize the protons against the head-tail instability. However during two stages of the squeeze when the beam separations were low, there were significant proton losses that were accompanied by a reduction in their bunch length. Particles with large momentum deviations were likely hitting synchro-betatron resonances and getting lost. Lowering the chromaticity to about 5 units still provided enough tune spread for stability but also lowered the proton losses and removed the longitudinal shaving [33].

A better-known phenomenon is the contribution of the interaction region to the chromaticity. At collision optics, the triplet quadrupoles contribute large linear and non-linear chromaticity as well as strong chromatic beta beats. The linear chromaticity is corrected to the desired value but the nonlinear chromatic effects, if not corrected, can lead to beam loss due to beam-beam or lattice driven synchro-betatron resonances. This was the experience in the Tevatron until 2006 when a second order chromaticity correction was put into effect [34]. This reduced the quadratic chromaticity by about a factor of five and decreased proton losses during stores.

*Experience at KEKB with chromatic coupling*

An interesting case of the combined effects of chromaticity and coupling has been recently reported from KEKB after the installation of crab cavities in 2007 in each ring. Coupling was found to be stronger for off-momentum particles both in measurements and simulations with their model lattice. Sources of this chromatic coupling were thought to be the misaligned sextupoles, higher order multiples in the final focus quadrupoles, special magnets and other lattice errors. Weak-strong and strong-strong beam-beam simulations showed that the luminosity was not sensitive to the chromatic coupling without the crab cavities but in their presence, the luminosity could drop as much as 10% due to chromatic effects [35].

KEKB operates close to the 1$^{st}$ order synchro-betatron resonance near the diagonal in tune space $q_x - q_y + q_s = N$, and various sources could be driving this resonance. Installation of skew sextupoles to control the chromatic coupling resulted in about 15% increase in luminosity [22]. Measurements showed that these skew sextupoles were effective in increasing the luminosity with the crab cavities turned off as well [23]. The maximum vertical beam-beam parameter achieved is 0.09 in the higher energy ring as opposed to a predicted value of 0.15 by beam-beam simulations. The reasons for the discrepancy and the limitations on achieving higher luminosity were under active study as of June 2010 [23].

It is an interesting question why KEKB was so susceptible to this chromatic coupling and not other accelerators such as PEP II, since rotational misalignments of sextupoles are not uncommon. It could simply be that KEKB operated closest to the linear synchro-betatron resonances. The tunes in PEP II appear to have been closest to the higher order resonance $q_x - q_y + 2 q_s = N$ in both rings [36] and may have therefore not been affected.

### 1.1.4.2    *Local coupling and dispersion*

In the Tevatron, global coupling is controlled to a minimum tune split of 0.002. Both in the Tevatron and in RHIC local decoupling in the Irs has been operational to correct for rotational misalignments of triplet quadrupoles (in some cases by several mrad) in order to optimize luminosity. Local dispersion is measured and corrected to within a few cm at the IPs in the Tevatron.

In lepton colliders there are direct geometrical effects since the coupling controls the vertical emittance and hence the vertical beam sizes at the IPs. KEKB finds it essential to correct both the local coupling and the dispersion at the IP during their luminosity optimisation.  They use anti-solenoids and skew quadrupoles to correct the local coupling sources and dipole correctors to correct the dispersions in both planes at the IP. In PEPII reducing the coupling in the interaction region of the low energy ring was found to be essential to increasing the luminosity. This was done by installing several permanent magnet skew quadrupoles in the IR [24]. It seems to be generally accepted that the dynamical effects of uncorrected coupling and dispersion have a greater impact on the luminosity than the purely geometrical effects in lepton colliders.

*1.1.4.3   Matching beam sizes*

SPS had reported that when proton emittances were 4 times larger than anti-proton emittances, protons could be lost due to high order resonances such as 13[th] and 16[th] order. From 1988 onwards, the emittance ratio was reduced to < 2, proton losses dropped as long as resonances of lower order such as the 10[th] were avoided. Dedicated studies were done to measure the impact of unequal emittances [29]. One proton (rms normalized emittance ~ 5.5 π mm-mrad) and one anti-proton bunch (rms normalized emittance ~ 7.5 π mm-mrad) were injected into the SPS and each collided twice with the other bunch per turn. The tunes were changed and proton background rates and lifetimes were measured first with the initial anti-proton emittances and then the anti-proton bunch was scraped to reduce its emittance to nearly equal the proton emittance and the loss rates measured again. In the first case with the larger and more intense anti-proton bunch, the proton bunch was not sensitive to 13[th] and 16[th] order resonances. In the second case with the smaller and less intense anti-proton bunch, the proton bunch was sensitive to these resonances even though the beam-beam parameter was about 40% lower. A scaling law such as the one in Equation (3) would not explain this dependence. A quantitative theoretical model to explain these observations has not yet been developed.

Observations in the Tevatron have been similar. When electron cooling of anti-protons in the Recycler made their emittance about 5-6 times smaller than those of protons, the latter suffered large losses [2]. A noise source was introduced to increase the anti-proton emittance and reduce the emittance ratio to about 3. This reduced the losses to acceptable levels.

HERA also had to control the mismatch but in their case, the beam sizes had to be matched to within 20% for tolerable beam losses [13]. This stringent tolerance is at first glance harder to understand given that the beam-beam parameter was about 0.001 compared to 0.005 in the SPS and about 0.008 in the Tevatron. One can speculate about possible reasons, e.g. the lower beam-beam spread allowed the proton tunes to lie closer to resonances bur made them more susceptible to small perturbations such as an increased non-linear field from the smaller opposing beam.

**1.1.5   Orbit vibrations at the IP**

Orbit vibrations at the IP modulate the offset between the colliding beams and are thought to lead to an emittance increase depending on the frequencies of modulation. Random orbit fluctuations at the IP have been theoretically shown to lead to diffusion and emittance growth [37].

Triplet vibrations in the frequency range from 4 to a few hundred Hz have been measured at the Tevatron and these frequencies have also been seen in the orbit spectrum [38]. Vibrations in this range are attributed to the liquid helium pumps, ground vibrations due to passing vehicles etc. An orbit feedback system installed in 2005 reduced the orbit drift during stores by a factor of eight and may have also helped to keep the bunches better centred at the IPs [39].

In RHIC orbit modulations such as those resulting from the 10 Hz vibrations of the triplet quadrupoles have long been thought to limit proton beam lifetime during stores [9]. Recent measurements showed that modulations of the betatron tunes and orbits could be well correlated with these vibrations [40]. It was suggested that the orbit modulation could manifest as a modulated crossing angle at the IPs and may explain the relative large proton losses at the start of stores in recent years.

In HERA, closed orbit oscillations of the electron beams were measured at the IPs with largest amplitudes at frequencies in the range 2-15 Hz. The sources were traced to vibrations of the electron triplet quadrupoles in the two IRs due to ground motion. These oscillations of the electron orbit led to increased proton background rates as the beams were brought into collision. A feedback system using BPMs upstream and downstream of the IPs was installed to control these oscillations [41].

### 1.1.6 Coherent phenomena

Coherent instabilities have long been observed in lepton colliders that operate with nearly equal intensities in both beams, see e.g. reference [42]. Observations of coherent beam-beam effects have been less frequent in hadron colliders. Beam loss due to coherent beam oscillations was reported in the ISR [17]. This usually occurred when the vertical separation between the beams was gradually reduced to initiate collisions. The losses started when the separations reached about $1\sigma$ and the beam-beam tune shift was about 0.001 per interaction region. The losses were reduced by a combination of reducing the separation at one interaction region at a time, improving the vertical feedback system and increasing the tune spread to increase Landau damping. SPS does not appear to have suffered from beam loss due to coherent oscillations, possibly due to the large difference in anti-proton and proton intensities.

In the Tevatron coherent instabilities do not cause beam loss during regular operation. There have been sporadic reports of multi-bunch coherent instabilities, usually when the chromaticity was too low [7]. However coherent dipole modes have been observed in recent dedicated studies [43]. The observed modes were in rough agreement with the coupled bunch mode spectrum calculated from a matrix analysis using 3 bunches per beam interacting only via the head-on interactions. However there were some observed frequencies that were unexpected.

RHIC reported the first observation of coherent modes in a hadron collider [44]. Both the sigma mode and the pi mode were observed during operation with protons beams with four head-on collisions per turn and a beam-beam parameter/IP of 0.0015. These modes, shown in Figure 8, appeared when the bunches were colliding and disappeared when the bunches were separated.

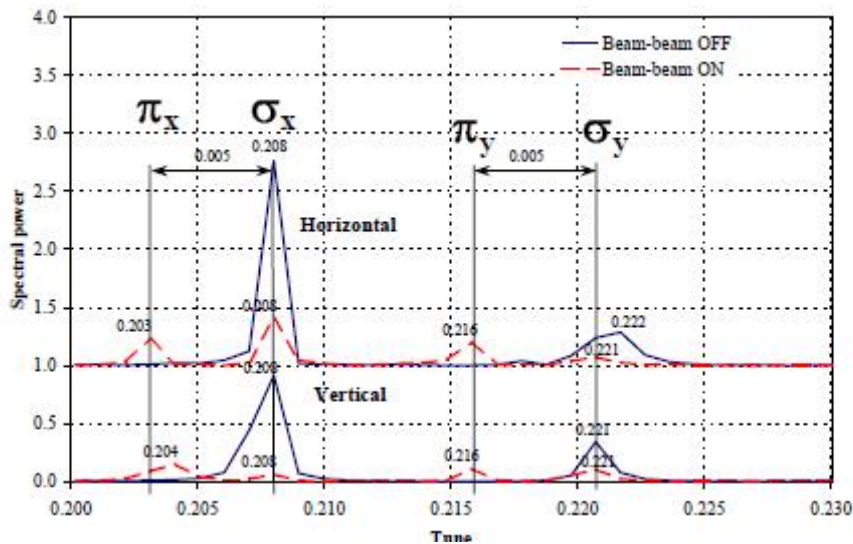

**Figure 8**: First observation of coherent beam-beam dipole modes in RHIC (taken from [44]).

They were also observed in a dedicated experiment with 1 collision per turn and beam-beam parameter = 0.003. These modes could be well reproduced in simulations [45]. More recent BTF measurements in 2009 have shown the appearance of sigma and pi modes in the vertical plane of both beams but not in the horizontal plane during regular operation [46]. No instability was associated with the appearance of these modes, especially the pi mode, which is outside the incoherent spectrum. This runs counter to theoretical expectations that the pi mode being undamped and would therefore, in the presence of machine impedance for example, initiate instabilities [47]. This needs to be better understood especially for the LHC where much effort has been put into understanding possible mechanisms for damping this mode, e.g. [48].

### 1.1.7 Compensation of head-on interactions with an electron lens

Compensation with an electron lens is covered elsewhere in this issue, so the discussion here will be brief. Operation with a Gaussian electron lens in the Tevatron has shown that it produces the expected tune shift and tune spread when acting on an anti-proton bunch [49]. Compensation of the head-on interactions has not yet been observed but simulations of the compensation in RHIC and the LHC has been done by three different codes with similar results [50, 51, 52]. They find the following
- The compensation works at higher values of the bunch intensity than at present used in operation in RHIC or the design value in LHC respectively. RHIC already suffers emittance growth and beam loss at present intensities. Even though the head-on collisions cause losses, the electron lens compensation does not become effective until higher intensities. What determines the critical bunch intensity above which the electron lens is useful?
- The electron lens intensity should not compensate more than half the tune spread due to the head-on interaction. At higher electron lens intensities and larger reduction of the tune spread, the proton beam lifetime suffers. Coherent instability due to a small tune spread is not the cause of this result since these were weak-strong simulations

- The electron transverse density should be uniform with a width larger than that of the proton bunch that is being compensated. With a wider lens, the proton bunch does not experience the sharp nonlinear fields at the edges pf the electron beam and effectively sees mostly the linear part of the force from the electron lens. However the effect of the electron lens is more beneficial than a simple tune shift.

These numerical predictions need to be tested with measurements. These will happen after electron lenses are installed in RHIC. If these predictions are borne out, then there is more to understand about the electron lens compensation.

### 1.1.8 Compensation of long-range interactions with wires

The principle of long-range compensation with a wire was partially tested in RHIC in 2009. The measurements were done in a single study where a single long-range interaction was created at a very small phase difference from the wire location [53]. Measurements of loss rates and bunch intensities showed that the wire reduced the losses for the beam in the Yellow ring but not for the beam in the Blue ring. Simulations seem to suggest that the separations between the beams (3.1σ) may not have been large enough for the wire compensation to be effective [54]. We recall the field due to the long-range interaction approaches the 1/r dependence of the field of a wire when the separations are significantly greater than 3σ. The wires have been removed from RHIC so further measurements may have to wait until wires are installed in the LHC during an upgrade. In earlier studies at RHIC, the effect of a wire on a beam was studied as a function of the beam-wire separation with different particle species at injection and collision [55]. Extensive simulations of the beam-wire interactions showed satisfactory agreement with the measurements [56]. The beam-wire distance at which the loss rates spiked found by simulations agreed to within 0.5σ with measurements at injection and collision and the higher loss rates observed with deuteron beams compared to gold beams were also reproduced in simulations.

### 1.1.9 Future developments related to the LHC

*Crab cavities*: Following the success with crab cavities in KEKB, there are plans to test the concept for implementation in the LHC during a future upgrade [57]. Two schemes are envisaged: a global scheme with a single cavity per ring or a local scheme with pairs of crab cavities around the high luminosity IRs. Some of the beam dynamics issues were examined in reference [58]. Some issues require detailed studies such as the sensitivity of the beam to phase noise in the cavities, synchro-betatron resonances driven by dispersion in these cavities and perhaps others.

*Crab waist*: The crab waist concept [59] has been demonstrated to work in DAΦNE [60]. The concept works for flat beams by placing sextupoles at appropriate phase advances in the IR such that the vertical phase advance in the IR becomes independent of horizontal betatron oscillations. This effectively suppresses some resonances driven by the beam-beam interactions. It is not immediately obvious that the same scheme will also work in hadron colliders with round beams where resonances with modulations of the horizontal phase are strong. Are there modifications of this scheme that can be successfully applied to hadron colliders?

*Flat bunches and large Piwinski angles*: One of the possible paths to higher luminosity at the LHC is the so-called large Piwinski angle (LPA) scheme in which bunches collide at an angle with a large Piwinski parameter ($\phi \sim 2$) and large bunch intensity keeping the beam-beam parameter at the same value as in other schemes [61]. The luminosity increases with the bunch intensity. The number of bunches is reduced to keep the beam current, hence the heat load, down. An additional 40% gain in luminosity is obtained if a longitudinally flat profile rather than a Gaussian profile is used. These bunch profiles have lower peak fields and hence lower electron cloud effects. Preliminary studies of beam-beam effects showed lower transverse diffusion than with Gaussian bunches [62]. This needs to be checked with more detailed studies. If these results are confirmed, this scheme with longitudinally flat profiles may be attractive even without large Piwinski parameters and high bunch intensities.

*Beam-beam limit at high energies*: There are plans to operate the LHC at more than double the design energy of 7 TeV. At such energies, effects of synchrotron radiation become much more important with the radiation damping time being of the order of an hour. Will the beam-beam limit be set by the saturation of a beam-beam parameter due to emittance growth (1$^{st}$ beam-beam limit in lepton colliders) or by the creation of tails and beam loss? This issue is already under study [20].

A general list of the beam-beam related issues in the LHC were discussed in reference [63]. Besides the effects discussed in this reference and those listed above, there are likely to be other manifestations of beam-beam effects at the LHC, some anticipated and some perhaps not. The multiple physics aspects of this effect will remain interesting in any circumstance.

.